\documentclass{elsart}
\usepackage{amssymb,amsmath}
\usepackage{epsf}
\def\LM#1#2{\left|\begin{array}{l}{#1}\\[1ex]{#2}\end{array}\right.}

\begin{document}

\begin{frontmatter}
\title{Subdiffusion-Limited Reactions}
\author{S. B. Yuste$^{1}$ and Katja Lindenberg$^{2}$}
\address{$^{1}$Departamento de F\'{\i}sica, Universidad  de  Extremadura,
E-06071 Badajoz, Spain\\
$^{2}$Department of Chemistry and Biochemistry 0340, and Institute
for Nonlinear Science,
University of California San Diego, La Jolla, California 92093-0340}

\begin{abstract}
We  consider the coagulation dynamics $A+A\rightarrow A$  and the
annihilation dynamics $A+A \rightarrow 0$
for particles moving subdiffusively in one dimension, both on a 
lattice and in a continuum.  The analysis combines the
``anomalous kinetics" and ``anomalous diffusion" problems, each
of which leads to interesting
dynamics separately and to even more interesting dynamics in
combination. We calculate both short-time and long-time concentrations, 
and compare and contrast the continuous and discrete cases.
Our analysis is based on the fractional diffusion equation and its discrete
analog.
\end{abstract}

\end{frontmatter}

\section{Introduction}

Diffusion-limited reactions in low-dimensional geometries
have been studied intensely because they exhibit ``anomalous kinetics,"
that is, behavior different from
that predicted by the laws of mass action in well-stirred
systems~\cite{book}.
Among the simplest and most extensively studied are
single species diffusion-limited coagulation ($A+A\rightarrow A$ or $A+A
\rightleftharpoons A$)~\cite{DBA} and annihilation ($A+A \rightarrow
0$)~\cite{DBA,ours}.  These
reactions, which show anomalous behavior in one dimension,
are of particular theoretical interest because they
lend themselves to {\em exact} solution in one
dimension.  Exact one-dimensional solutions can be obtained for
diffusion in 
continuous~\cite{DBA,TorMcCJPC,Spouge,AvrBurDoeJSP,AbraMPL,MasAvrPLA,wenew}
as well as
in discrete~\cite{DBA,KrebsEtAl,HenHin,AbaFriNic} systems.
A particularly elegant solution of the coagulation problem is provided
by the method of intervals.  This method focuses on the diffusive
evolution of empty intervals, that is, of intervals that contain no
particles~\cite{DBA,AvrBurDoeJSP,AbraMPL,MasAvrPLA,ManAvr}. The empty interval
equation turns out to be linear and hence exactly solvable.  Originally
developed for continuous systems, the method has been extended to discrete
lattices~\cite{KrebsEtAl,HenHin,AbaFriNic}.  The method of intervals can
not be directly adapted to the annihilation reaction, but a new formalism,
the method of odd/even intervals which keeps track of the parity (even or
odd) of the number of particles in an interval, has recently been
developed~\cite{MasAvrPLA,wenew}.  Again, the odd/even interval equation
turns out to be linear and hence exactly solvable.

The anomalies of the $A+A$ problem in one dimension are typically
displayed in two ways:
one is through the time dependence of the reactant
concentration $c(t)$, which for the
$A+A\rightarrow A$ and the $A+A\rightarrow 0$ reactions in infinite systems
decays asymptotically as $t^{-1/2}$ instead of the law of mass
action decay $t^{-1}$.  The other is through the interparticle
distribution function
$p(x,t)$, which is the (conditional) probability density for
finding the nearest particle at a distance $x$ on one side of a given
particle.  This function scales as $x/t^{1/2}$, in typical diffusive
fashion.  In one dimension a gap develops around each particle
that leads to a more ordered spatial distribution than the
exponential distribution implicit in well-stirred systems and ``explains"
the relative slowing down of the reaction.

In a parallel development, the problem of ``anomalous diffusion" has also
attracted a great deal of
attention~\cite{Kehr,HilferEd,MetKlaPhysRep}.
The universally accepted characterization of
anomalous (as in ``not ordinary") diffusion
is through
the mean squared displacement of a process $x(t)$ for large $t$,
\begin{equation}
\left< x^2(t)\right> \sim \frac{2K_\alpha}{\Gamma(1+\alpha)} t^\alpha .
\label{meansquaredispl}
\end{equation}
Ordinary diffusion ($\alpha=1$, $K_1\equiv D$) follows Gaussian
statistics and Fick's second law for the probability density for finding
the process at $x$ at time $t$,
\begin{equation}
\frac{\partial}{\partial t} P(x,t)= D \frac{\partial^2}{\partial x^2}
P(x,t),
\label{Peqdifun}
\end{equation}
leading to linear growth of $\left< x^2(t)\right>$ with time. 
Anomalous diffusion is characterized
by a nonlinear dependence.
If $0<\alpha<1$ the process is subdiffusive or
dispersive; if
$\alpha>1$ it is superdiffusive.  Anomalous diffusion is associated with
many physical systems and is not due to any single universal cause, but
it is certainly ubiquitous.   Nor is anomalous diffusion modeled in a
universal way; among the more successful approaches to the subdiffusive
problem, which is the case we consider in this work,
have been continuous time random walks with non-Poissonian
waiting time distributions~\cite{Kehr}, and fractional
dynamics approaches in which the diffusion equation (\ref{Peqdifun})
is replaced by the generalized diffusion
equation~\cite{HilferEd,MetKlaPhysRep,Balakrishnan,SchWysJMP,DoeAvrPRL,MetBarKlaPRL,MetBarKlaEPL}
\begin{equation}
\frac{\partial }{\partial t} P(x,t)=
K_\alpha ~_{0}\,D_{t}^{1-\alpha } \frac{\partial^2}{\partial x^2} P(x,t)
\label{Pfracdifu}
\end{equation}
where  $~_{0}\,D_{t}^{1-\alpha } $ is the Riemann-Liouville operator,
\begin{equation}
~_{0}\,D_{t}^{1-\alpha } P(x,t)=\frac{1}{\Gamma(\alpha)}
\frac{\partial}{\partial t} \int_0^t d\tau
\frac{P(x,\tau)}{(t-\tau)^{1-\alpha}},
\end{equation}
and $K_\alpha$ is the generalized diffusion coefficient
that appears in Eq.~(\ref{meansquaredispl}).
Some limitations of this description and some connections between the
generalized diffusion equation and continuous random walk formulations
have been discussed recently~\cite{BarMetKlaPRE,SokBluKlacondmat}.

In this paper we consider a combination of these two phenomena, namely,
the kinetics of $A+A$ reactions of particles that
move subdiffusively in one dimension~\cite{ourPRL}.
Some aspects of this problem have also been considered using the waiting time
distribution approach~\cite{Blumen,AlemanyJPA}.  Those solutions require
approximations relating the reactant concentration to the distinct
number of sites visited by a particle~\cite{Blumen}, or the
waiting time distributions for single particles to the waiting time
distributions for relative motion~\cite{AlemanyJPA}.
Here we adapt the
fractional dynamics approach to the problem and take advantage of the fact
that the resulting generalized diffusion equations can
be solved in closed form~\cite{ourPRL}.  We consider both
coagulation and annihilation reactions.  

In Section~\ref{evolution} we develop the fractional diffusion
equations from which one obtains the concentration of reactant as a
function of time.  For the coagulation reaction we generalize the empty
interval method~\cite{DBA,AvrBurDoeJSP} and, for the annihilation
reaction, the odd/even interval method~\cite{MasAvrPLA,wenew}. 
The solutions are presented in Section~\ref{solanni}.
Sections~\ref{evolution} and \ref{solanni} deal with kinetics on 
continuum one-dimensional systems; in Section~\ref{lattice} we extend
the analysis
to reactions on a lattice.  The differences between the continuum and
discrete results are most pronounced at early times (long-time differences
are due to finite size effects that get pushed to longer times if one deals
with larger systems).  We conclude with a recapitulation in
Section~\ref{conclusions}.

\section{Evolution equations for subdiffusive particles}
\label{evolution}
\subsection{Empty interval method for coagulation reaction}

Consider first the coagulation reaction $A+A \rightarrow A$ when the
particles move by ordinary diffusion. The probability distribution
function for the position $x$ of any {\em one} $A$ particle {\em in
the absence of reaction} obeys the diffusion equation (\ref{Peqdifun}).
The coagulation problem can be formulated in terms of the probability 
$E(x,t)$ that an interval of length $x$ is empty of particles
at time $t$.  This ``empty interval" function obeys the 
diffusion equation~\cite{DBA,AvrBurDoeJSP}
\begin{equation}
\frac{\partial}{\partial t} E(x,t)= 2D \frac{\partial^2}{\partial x^2} E(x,t).
\label{Eeqdifun}
\end{equation}
The derivation of this equation is
straightforward and recognizes that an empty interval is shortened
or lengthened by movement of particles in and out at either end
according to the dynamics described by Eq.~(\ref{Peqdifun}).

It is instructive to recall this derivation as given in
Refs.\ \cite{AvrBurDoeJSP,AbraMPL}.  The
reasoning is more transparent if the diffusion processes that
lead to the change $\partial E(x,t)/\partial t$ are described on a
lattice with lattice spacing $\Delta x$ and sites labeled by integers.
In this lattice $E_n(t)$ gives the probability that sites $1$ through $n$, for
example, are empty at time $t$.  Since the event $E_{n+1}$ contains the
event $E_n$,
the probability that sites $1$ through $n$ are empty but that there is a
particle at site $n+1$ is $E_{n+1}-E_n$. 
The particles move randomly to the nearest lattice site with a
hopping rate $2D/(\Delta x)^2$, the rate
$D/(\Delta x)^2$ to the right being equal to the rate to the left. 
If the particle
at $n+1$ moves to site $n$ in a short time interval $\Delta t$, $E_n$ will
decrease by 
$[D/(\Delta x)^2](E_{n}-E_{n+1})\Delta t$. On the other hand, $E_n$ may
increase by the departure of a particle at site $n$ to site $n+1$; 
the associated increase is $[D/(\Delta x)^2](E_{n-1}-E_{n})\Delta t$.
The same entry and exit processes can take place at the other end of the
interval.  Combining these four processes then gives for the change in
$E_n$
\begin{equation}
\frac{\Delta E_n}{\Delta t} = \frac{2D}{(\Delta x)^2} (E_{n+1} -2E_n
+E_{n-1}).
\label{En}
\end{equation}
The coagulation reaction fixes the boundary condition.  Coagulation
involves adjacent occupied sites and either of the two particles hopping
onto the other, thereby disappearing.  Two adjacent sites can both
be occupied, or one can be occupied and the other empty (probability
$2(E_1-E_2)$ because this can occur in two ways) or both can be
empty (probability $E_2$).  Since this covers all possibilities, the
probability that two adjacent sites are occupied is therefore $1-2E_1+E_2$.
The change in $E_1$ is thus 
\begin{equation}
\frac{\Delta E_1}{\Delta t} = \frac{2D}{(\Delta x)^ 2} (1-2E_1+E_2),
\label{E1}
\end{equation}
and for this equation to fit the pattern of Eq.~(\ref{En}) it is necessary
to require the boundary condition 
\begin{equation}
E_0=0.
\label{E0}
\end{equation}
Letting $x=n\Delta x$ and both $\Delta x$ and $\Delta t \rightarrow 0$ then
leads to the continuous equation (\ref{Eeqdifun}) with the boundary
condition
\begin{equation}
E(0,t)=0.
\label{E0c}
\end{equation}
The other boundary condition requires that the population of particles be
nonvanishing: 
\begin{equation}
E(\infty,t)=0.
\label{Einf}
\end{equation}
The empty interval dynamics is thus
essentially the same as that of individual particles in the absence of
reaction, but with different boundary conditions and a diffusion coefficient
$2D$ that reflects the fact that
the relative motion of two diffusive particles involves twice the
diffusion coefficient of each particle alone.

The connection between the empty interval function and the observables of
interest is obtained as follows~\cite{AvrBurDoeJSP,AbraMPL}.  
Back in the discrete formulation, the
probability that a site is occupied (i. e. not empty) is $1-E_1$, and the
concentration of particles is therefore 
\begin{equation}
c(t) = \frac{(1-E_1)}{\Delta x} ~~~\longrightarrow~~~ c(t) = - \left.
\frac{\partial E(x,t)}{\partial x}\right|_{x=0},
\label{c}
\end{equation}
where we have exhibited the continuum limit.  The interparticle
distribution function $p_n(t)$ is the probability that the nearest neighbor
to one side of a given particle is $n$ lattice spacings away.  This
probability is related to $E_n$ by
\begin{equation}
E_n = c\Delta x \sum_{k=n+1}^{\infty} \sum_{m=k}^{\infty} p_m.
\label{dipdf}
\end{equation}
The first sum insures that the next neighbor is at least $k$ sites away and
the second that $k$ is at least $n+1$.  The normalization takes into
account that the average distance between the particles is the reciprocal
of the concentration,
\begin{equation}
\left< n\Delta x\right> = \sum_{n=1}^{\infty} np_n\Delta x =\frac{1}{c}.
\end{equation}
Equation~(\ref{dipdf}) can be inverted:
\begin{equation} 
c\;p_n = \frac {(E_{n+1}-2E_n+E_{n-1})}{\Delta x} ~~~ \longrightarrow ~~~
c(t)\;p(x,t) =\frac{\partial ^2E(x,t)}{\partial x^2}
\label{ipdf}
\end{equation}
where again we have exhibited the continuum limit.

The construction of the kinetic equation for $E(x,t)$ for subdiffusive
particles proceeds along arguments analogous to those used in
the diffusive case.  Again, one follows the motion of the particles in
and out at the ends of the empty interval according to the dynamics
(\ref{Pfracdifu}). 
However, for particles that move subdiffusively, the
concept of a hopping rate (steps per unit time) is not defined because
the number of steps
performed up to time $t$ by a subdiffusive particle goes as
$t^{\alpha}$ with $\alpha<1$~\cite{SokBluKlacondmat}.
Therefore, the derivation of the empty interval equation must be
adjusted accordingly.  Proceeding as above in the discrete formulation,
we see that the probability that sites $1$ through $n$ are empty but that
there is a particle at site $n+1$ is still $E_{n+1}-E_n$.  However, the
rate of decrease of $E_n$ due to a particle that moves from site $n+1$ to
site $n$ is now described by the generalized diffusion process implicit in 
Eq.~(\ref{Pfracdifu}), that is, $[K_\alpha
~_{0}D_t^{1-\alpha}/(\Delta x)^2](E_n-E_{n+1})\Delta t$.  Collecting
arrival and
departure processes at both ends of the interval then leads to the equation
\begin{equation}
\frac{\Delta E_n}{\Delta t} = \frac{2K_\alpha ~_{0}D_t^{1-\alpha}}{(\Delta
x)^2}
(E_{n+1} -2E_n +E_{n-1}).
\label{Ensub}
\end{equation}
In the continuum limit this then becomes
\begin{equation}
\frac{\partial }{\partial t} E(x,t)=
2K_\alpha ~_{0}\,D_{t}^{1-\alpha } \frac{\partial^2}{\partial x^2} E(x,t).
\label{Efracdifu}
\end{equation}
The boundary conditions are exactly as before, Eqs.~(\ref{E0c}) and
(\ref{Einf}), as are the relations (\ref{c}) and (\ref{ipdf}) to the
observables.

\subsection{Odd/even interval method for annihilation reaction}
The empty interval method 
can not be applied to the annihilation reaction $A+A\rightarrow 0$
because annihilation leads to a discontinuous growth of empty intervals.
However, recently a new method of odd/even intervals has been introduced 
that leads to exact solution in the diffusion-limited case.  It 
is based on the construction of an equation for $r(x,t)$,
the probability that an arbitrary interval of length
$x$ contains an even number of particles at time
$t$~\cite{MasAvrPLA,wenew}. This probability does not change if two
particles inside the interval react since this process does not change the
even/odd parity of the interval.  
Because $r(x,t)$ changes only by the movement of particles in or  
out of the ends of the interval, arguments similar to those that lead to
Eq.~(\ref{Eeqdifun}) lead to the same equation for $r(x,t)$. 
This procedure can again be directly extended to the
subdiffusive problem, and $r(x,t)$ satisfies the same
fractional diffusion equation as $E(x,t)$:
\begin{equation}
\frac{\partial }{\partial t} r(x,t)= 2 K_\alpha
~_{0}D_{t}^{1-\alpha }  \frac{\partial^2}{\partial x^2} r(x,t) .
\label{rbasica}
\end{equation}
One boundary condition is the same as for $E(x,t)$ and is obtained via the
same arguments used earlier:
\begin{equation}
r(0,t)=1.
\label{r1}
\end{equation}
The other boundary condition in general differs from that of the empty
interval method and depends on the initial condition. In
particular, it is determined by the parity of total initial number of
particles.  This parity never changes.
A random initial placement leads to an equal probability that the system
forever contain an even or and odd number of particles, so that
\begin{equation}
r(\infty,t)=\frac{1}{2}.
\label{rccxinfini}
\end{equation}
The concentration of particles is related to
$r(x,t)$ precisely as in Eq.~(\ref{c}):
\begin{equation}
c(t)=-\left. \frac{\partial r(x,t)}{\partial x}\right|_{x=0}.
\label{cr}
\end{equation}

\section{Solution of the fractional diffusion equations and subdiffusive
reaction kinetics}
\label{solanni}
\subsection{Coagulation reaction}

The solution of Eq.~(\ref{Eeqdifun}) with boundary conditions
(\ref{E0c}) and (\ref{Einf}) can be written as~\cite{SchWysJMP}
 \begin{equation}
E(x,t) =\int_0^\infty dy \left[W(x-y,t) -W(-x-y,t)\right]
E(y,0)+
       \int_0^{t(x/\sqrt{2 K_\alpha})^{-2/\alpha}}
        \omega_{\alpha/2}(\eta) d\eta,
       \label{EsolSW}
\end{equation}
where
\begin{equation}
W(x,t)=\frac{1}{\sqrt{8 K_\alpha t^\alpha}}
H^{10}_{11}\left[\frac{|x|}{\sqrt{2K_\alpha  t^{\alpha}} }
	\LM{(1 -\frac{\alpha}{2},\frac{\alpha}{2})}{(0,1)}   \right],
\end{equation}
\begin{equation}
\omega_\beta(x)=\frac{1}{\beta x^2}
H^{10}_{11}\left[\frac{1}{x }
	\LM{(-1,1)}{(-1/\beta,1/\beta)}   \right],
\end{equation}
and  $H$ is the Fox
H-function~\cite{HilferEd,MetKlaPhysRep,SchWysJMP,MathaiSaxena}.
Taking into account that in Laplace transform space (indicated by a tilde over the function and
the letter $u$ as variable)
\begin{equation}
\widetilde W(x,u)=\frac{u^{\alpha/2-1}}{\sqrt{8 K_\alpha}}
\exp\left[ \frac{u^{\alpha/2}}{\sqrt{2K_\alpha}} |x| \right]
\end{equation}
and $\widetilde \omega_\beta(u)=\exp(-u^\beta)$,  the solution $E(x,t)$ in
Laplace space adopts a much simpler form:
\begin{equation}
\widetilde{E}(x,u)=\frac{s}{2 u}
\int_0^\infty dy \left( e^{-|x-y| s }  -  e^{-|x+y|s } \right)   E(y,0)  
+ \frac{1}{u}   \exp\left(-x s \right)
\label{EsolLap}
\end{equation}
where 
\begin{equation}
s\equiv \frac{u^{\alpha/2}}{\sqrt{2K_\alpha}}.
\label{sdefi}
\end{equation}
From Eqs.~(\ref{c}) and (\ref{EsolLap}) one finds
\begin{equation}
\widetilde c(u)=\frac{s}{u}
\left[1-s \widehat E\left(s,0 \right)\right] ,
\label{cuE}
\end{equation}
where $\widehat E(v,t)=\int_0^\infty dx e^{-v x} E(x,t)$
is the the {\em spatial}
Laplace transform of $E(x,t)$.
Equivalently, from Eq.\ (\ref{ipdf}) one obtains
\begin{equation}
\widetilde c(u)=\frac{\lambda}{u}
 \left[1-\widehat p\left(s,0 \right) \right]
\label{cu}
\end{equation}
where $\lambda\equiv c(0)$ and $\widehat p(u,0)$  is the {\em spatial}
Laplace transform of the initial interparticle distribution function
$p(x,0)$.

A commonly considered initial interparticle distribution is the random
(Poisson) distribution of average concentration $\lambda$, 
for which $E(x,0)=e^{-\lambda x}$ and   
$p_0(x)=\lambda e^{-\lambda x}$.  For this initial distribution
either Eq.~(\ref{cuE}) or Eq.~(\ref{cu}) can be used to obtain
\begin{equation}
\widetilde c(u) =\frac{\lambda}{u+\lambda \sqrt{2K_\alpha} u^{1-\alpha/2}}
\end{equation}
and $c(t)$ is given in closed form in terms of
the Mittag-Leffler function~\cite{MetKlaPhysRep,PodlubMainar} of
parameter $\alpha/2$:
\begin{equation}
c(t)= \lambda E_{\alpha/2}   \left(-\lambda \sqrt{2K_\alpha} t^{\alpha/2}
\right)
\label{ctRandCoagu}
\end{equation}
The Mittag-Leffler function can be calculated by the series
expansion~\cite{MetKlaPhysRep,PodlubMainar}
\begin{equation}
E_\alpha(z)=\sum_ {k=0}^\infty \frac{z^k}{\Gamma(k \alpha +1)}.
\label{MittagLeffler}
\end{equation}
When $\alpha=1$ one recovers the usual result for
diffusion-limited coagulation~\cite{TorMcCJPC,Spouge} since the
Mittag-Leffler function of parameter $1/2$ is
$E_{1/2}(-x)=\exp(x^2)\text{erfc}(x)$.

Another initial particle distribution of interest is an ordered or periodic
arrangement,
$p(x,0)=\delta(x-1/\lambda)$.  In this
case Eq.~(\ref{cu}) leads to
\begin{equation}
\widetilde c(u) =\frac{\lambda}{u} \left[1-\exp\left(-\frac{u^{\alpha/2}}{\lambda \sqrt{2K_\alpha}}
\right)\right]
\label{cu2}
\end{equation}
and $c(t)$ is given in terms of the Fox H-function $H^{1,0}_{1,1}$ \cite{SchWysJMP,MathaiSaxena,AbraMPL}:
\begin{equation}
\frac{c(t)}{\lambda} =
1-
H^{10}_{11}\left[\frac{1}{\lambda \sqrt{2K_\alpha  t^{\alpha}} }
	\LM{(1,\frac{\alpha}{2})}{(0,1)}   \right].
\label{ctPeriodica}
\end{equation}
For normal ($\alpha=1$) diffusive particles one recovers the usual result for diffusion-limited coagulation~\cite{Spouge} since
$H^{10}_{11}\left[z
	\LM{(1,\frac{1}{2})}{(0,1)}   \right]=\text{erfc}\left(\frac{z}{2}\right) $.

It is useful to exhibit explicitly the long-time and short-time behavior of
the particle concentration.  The long-time behavior can be extracted via
Tauberian theorems from the small-$u$ behavior {\em independent of initial
condition} and therefore in particular common to both of the
initial conditions considered above,
\begin{equation}
\widetilde c(u)\sim \frac{u^{-\alpha/2-1}}{\sqrt{2K_\alpha}},
\label{cuasin}
\end{equation}
from which it follows that
\begin{equation}
c(t)\sim \frac{t^{-\alpha/2}}{\sqrt{2 K_\alpha}\Gamma\left(1-\frac{\alpha}{2}\right)}         .
\label{ctasinCoagu}
\end{equation}
For diffusive particles ($\alpha=1$) one recovers the well-known result 
$c(t)\sim t^{-1/2}$.
The short-time behavior of the particle concentration is, not surprisingly,
markedly different for the two initial conditions.  For the random initial
condition an expansion of Eq.~(\ref{ctRandCoagu}) using
(\ref{MittagLeffler}) gives
\begin{equation}
\frac{c(t)}{\lambda}=1-\frac{\lambda
\sqrt{2 K_\alpha}}{\Gamma(1+\frac{\alpha}{2})} t^{\alpha/2}+\ldots
\label{ct0Rand}
\end{equation}
Notice that this implies a infinite initial reaction rate because
$dc(t)/dt\rightarrow \infty$ for
$t\rightarrow 0$, a reflection of the fact that a random distribution of
particles includes a large probability of proximate
particles.  On the other hand, 
using the  asymptotic expression of  the Fox H-function $H^{1,0}_{1,1}$ for
large argument~\cite{SchWysJMP,MathaiSaxena} in the solution for the
periodic initial distribution of particles, Eq.~(\ref{ctPeriodica}), one
finds
\begin{equation}
\label{ct0Perio}
\frac{c(t)}{\lambda}=1-
\frac{1}{\sqrt{\pi (2-\alpha)}}
\left(  \frac{\alpha}{2}\right)^{1/(\alpha-2)}
 z ^{1/(\alpha-2)}
\exp\left[-\frac{2-\alpha}{2} \left(  \frac{\alpha}{2}\right)^{\alpha/(2-\alpha)} z^{2/(2-\alpha)}
\right] +\ldots
 \end{equation}
where   $z=[\lambda (2K_\alpha  t^{\alpha})^{1/2} ]^{-1}  $.
In this case $dc(t)/dt\rightarrow 0$ for $t\rightarrow 0$, since initially
there are no proximate particles.

\begin{figure}
\begin{center}
\leavevmode
\epsffile{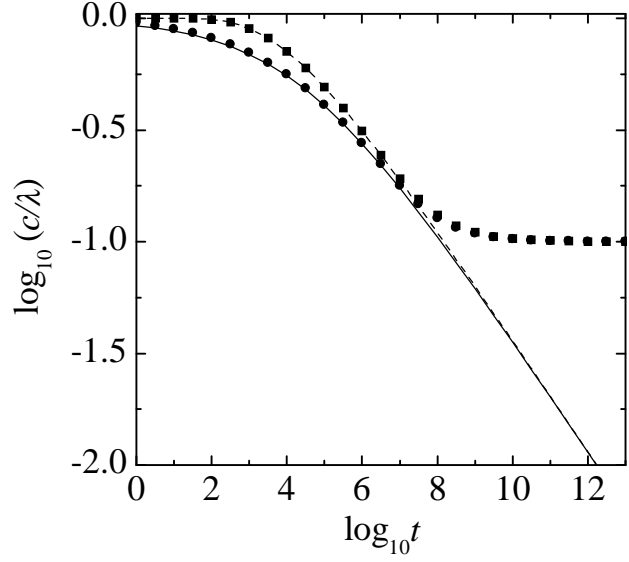}
\end{center}   
\caption{
Logarithm of the survival probability $c(t)/\lambda$  for 
coagulation dynamics of subdiffusive particles with $\alpha=1/2$,
$K_\alpha=1$, and initial concentration $\lambda=1/20$, versus the
logarithm of time.  The initial arrangements  of the
particles are Poissonian (solid line and circles) and periodic (broken
line and squares). The lines are the continuum solutions 
given by Eqs.~(\ref{ctRandCoagu}) and (\ref{ctPeriodica}).
The symbols are the on-lattice solutions for a finite lattice with
$L=200$ and $\Delta x=1$ with periodic boundary conditions.
} 
\label{fig:ctcoagu}
\end{figure}

The curves in Fig.~\ref{fig:ctcoagu} shows the time evolution of the
concentration $c(t)$ for the coagulation reaction and the two initial
conditions considered here.  The differences at early times are evident, as
is the coagulation of the curves into a single asymptotic form at long
times.

\begin{figure}
\begin{center}
\leavevmode
\epsffile{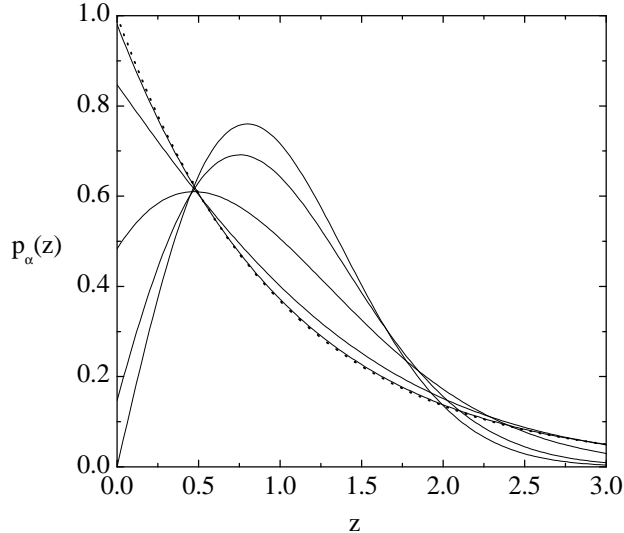}
\end{center}   
\caption{Long-time scaled interparticle distribution function for the
coagulation reaction
for several values of the anomalous diffusion exponent.  Proceeding
upward from lowest to highest curves along the $y$ axis intersection:
$\alpha=1, 0.95, 0.8, 0.5, 0.2$.  
Note that the distribution for $\alpha=0.2$ on this scale is nearly
indistinguishable from the completely random distribution $\exp(-z)$
(dotted curve).
} 
\label{fig:IPDF}
\end{figure}

The time-asymptotic form of the empty interval function can be used to
obtain the asymptotic behavior of the interparticle distribution function
via Eq.~(\ref{ipdf}).  From Eq.~(\ref{EsolLap}) one deduces that
\begin{equation}
\frac{\partial ^2}{\partial x^2}\widetilde{E}(x,u)=\frac{s^3}{2 u}
\int_0^\infty dy \left( e^{-|x-y| s }  -  e^{-|x+y|s } \right)   E(y,0)
+ \frac{s^2}{u}  e^{-x s},
\label{ParLapE}
\end{equation}
which for $u\rightarrow 0$ reduces to
\begin{equation}
\frac{\partial^2}{\partial x^2} \widetilde  E(x,u)\sim \frac{s^2}{u}  e^{-x s}.
\label{d2ELapasin}
\end{equation}
From Eq.~(\ref{ipdf}) and the fact that the inverse transform of
(\ref{d2ELapasin}) can be expressed in terms of a Fox H-function, one finds 
\begin{equation}
c(t)\; p(x,t)\sim  \frac{1}{x^2}
H^{10}_{11}\left[\frac{x}{\sqrt{2K_\alpha} t^{\alpha/2}}
	\LM{(1 ,\frac{\alpha}{2})}{(2,1)}   \right],
\label{Extasin}
\end{equation}
which in combination with Eq.~(\ref{ctasinCoagu}) immediately leads to
\begin{equation}
p(x,t) \sim \frac{\Gamma(1-\frac{\alpha}{2})
}{\sqrt{2K_\alpha}t^{\alpha/2}} H^{10}_{11}
\left[\frac{x}{\sqrt{2K_\alpha} t^{\alpha/2}}
	\LM{(1-\alpha ,\frac{\alpha}{2})}{(0,1)}   \right]
\label{pxttasin}
\end{equation}
for $t\rightarrow \infty$.  This result is independent of the initial
distribution of particles.
We can also write this in the scaled form
\begin{equation}
p_\alpha(z) \sim \Gamma^2\left(1-\frac{\alpha}{2}\right)
H^{10}_{11}\left[\Gamma\left(1-\frac{\alpha}{2}\right)z
\LM{\left(1-\alpha ,\frac{\alpha}{2}\right)}{(0,1)}   \right]
\label{pztasin}
\end{equation}
where $z\equiv c(t)x$ is the scaled interparticle distance
and $p_\alpha(z)dz\equiv p(x,t)dx$.
This stationary form is shown in Fig.~\ref{fig:IPDF}
for  several values of the diffusion exponent $\alpha$.

The interparticle distribution function conveys the interesting
``anomalies" of the problem most clearly.  For a random distribution of
particles on a line this distribution is exponential.  In particular,
the most probable interparticle gaps are the smallest.
For diffusion-limited reactions on a line it is well known that 
the scaled distribution deviates
in two ways from the exponential behavior. First, a gap develops around
each particle, and the distribution vanishes near the origin (see
$\alpha=1$ curve in the figure), indicating an ``effective repulsion"
of particles.  Second, the probability of large gaps decays much more
rapidly than exponentially: the decay goes as a power of $\exp(-z^2/2)$.  
In the subdiffusive case decreasing $\alpha$ leads to the diminution 
of the gap around each particle, that is, to a weakening of the
effective repulsion and to a behavior that appears closer to that of
a random distribution in the short-interparticle-distance behavior.
This is evident in the progression of the curves with decreasing
$\alpha$ shown in the figure.  Furthermore, 
the probability of large gaps decays as a power of
$\exp(-x^{2/(2-\alpha)})$, thus neatly interpolating 
between the purely random exponential decay 
state as $\alpha\rightarrow 0$ (since then
$p(x,t)\rightarrow c(t) e^{-c(t) x}$)
and the more ordered state corresponding to diffusive particles at
$\alpha=1$.

\subsection{Annihilation reaction}

It is straightforward to deduce that the solution $r(x,t)$ of
Eq.~(\ref{rbasica}) with initial condition $r(x,0)=\frac{1}{2}+E(x,0)/2$
and boundary conditions (\ref{r1}) and 
(\ref{rccxinfini}) is simply $r(x,t)=\frac{1}{2}+E(x,t)/2$ where $E(x,t)$
is the solution found in the previous section.  We can therefore here rewrite
Eq.~(\ref{cuE}) for the Laplace transform of the concentration of particle
undergoing the annihilation reaction as
\begin{equation}
\widetilde c(u)=\frac{s}{u}
\left[1-s \widehat r\left(s,0 \right)\right]
\label{ruE}
\end{equation}
where $\widehat r(v,t)$ is the the {\em spatial} Laplace transform
of $r(x,t)$. 

For a random initial distribution of mean concentration
$\lambda$ the initial distribution is
$r(x,0)=\frac{1}{2}+\frac{1}{2}e^{-2\lambda x}$ (note that this goes to
$1/2$ as $x\rightarrow\infty$ since there is an equal probability that we
start with en even or odd number of particles, and it goes to $0$ as
$x\rightarrow 1$ since zero particles is an even number).  Hence
\begin{equation}
\widetilde c(u) =\frac{\lambda}{u+2\lambda \sqrt{2K_\alpha}
u^{1-\alpha/2}},
\end{equation}
which can be inverted analytically to yield
\begin{equation}
c(t)= \lambda E_{\alpha/2}   \left(-2 \lambda \sqrt{2K_\alpha} t^{\alpha/2}
\right) .
\label{ctRandAniqui}
\end{equation}
Comparing this result with Eq.~(\ref{ctRandCoagu}) for the
coagulation reaction we see
that the rate of change of the concentration for annihilation is exactly
twice that for coagulation at all times, a result that has been noted
for ordinary diffusion~\cite{MasAvrPLA,wenew}.

For a periodic initial distribution with separation $1/\lambda$ between
particles we have 
\begin{equation}
r(x,0)=
\begin{cases}
2j+1-\lambda x & 2j/\lambda\leq x\leq (2j+1)/\lambda \\[9pt]
-2j-1+\lambda x & (2j+1)/\lambda\leq x\leq (2j+2)/\lambda 
\end{cases}
\end{equation}
This is a periodic function of period $2/\lambda$, so that~\cite{abramo}
\begin{equation}
\widehat r(u,0) =\dfrac{\int_0^{2/\lambda} dx e^{-u x}.
r(x,0)}{1-e^{-2u/\lambda}} 
\end{equation}
Carrying out this integral and substituting into Eq.~(\ref{ruE})
immediately leads to
\begin{equation}
\widetilde c(u) =
\frac{\lambda}{u}
\tanh \left( \frac{s}{2\lambda} \right)
= \frac{\lambda}{u}
\left[
\frac{
1-\exp\left(-\frac{u^{\alpha/2}}{\lambda \sqrt{2K_\alpha}} \right)}
{
1+\exp\left(-\frac{u^{\alpha/2}}{\lambda \sqrt{2K_\alpha}} \right)}
\right]
\label{cuAniquiPeriod}
\end{equation}
where we have written the second equality to make comparison with
Eq.~(\ref{cu2}) more apparent.  We have not been able to invert this
expression analytically to find $c(t)$ in closed form.

\begin{figure}
\begin{center}
\leavevmode
\epsffile{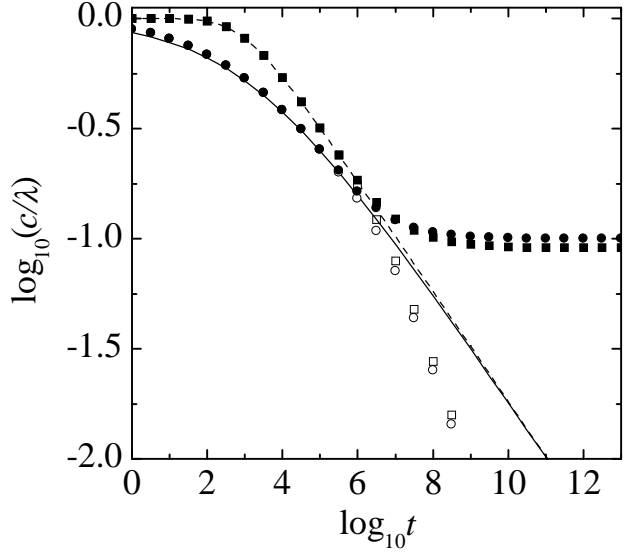}
\end{center}   
\caption{
Logarithm of the survival probability $c(t)/\lambda$  for
annihilation dynamics of subdiffusive particles with $\alpha=1/2$,
$K_\alpha=1$, and initial concentration $\lambda=1/20$, versus the
logarithm of time.  The initial arrangements  of the
particles are Poissonian (solid line and circles) and periodic (broken
line and squares). The lines are the continuum solutions
given by Eq.~(\ref{ctRandAniqui}) and by the numerical inversion of
Eq.~(\ref{cuAniquiPeriod}).
The symbols are the on-lattice solutions for a finite lattice with
$L=220$ (filled symbols) and $L=200$ (open symbols) with periodic boundary
conditions and $\Delta x=1$.  Note that with
these values of $L$ and $\lambda$ the initial number of particles and hence
the total number of particles is odd (filled symbols) or even (open
symbols) at all times. 
}
\label{fig:ctaniqui}
\end{figure}

Figure~\ref{fig:ctaniqui} shows the temporal evolution of $c(t)$ for the
two initial conditions when $\alpha=1/2$.  The evolution for a periodic
initial distribution has been obtained by numerically inverting
Eq.~(\ref{cuAniquiPeriod}).

The long-time behavior is again seen to be independent of initial condition
and can be obtained directly from Eq.~(\ref{ruE}).
Since $\widehat r(s,0)\rightarrow (2s)^{-1}$ when $s\rightarrow 0$, it
follows that as $u \rightarrow 0$,
\begin{equation}
\widetilde c(u)\sim \frac{u^{-\alpha/2-1}}{2\sqrt{2K_\alpha}}
\label{cuasinAniqui}
\end{equation}
which in turn implies that 
\begin{equation}
c(t)\sim \frac{t^{-\alpha/2}}{2\sqrt{2 K_\alpha}\Gamma\left(1-\frac{\alpha}{2}\right)},           
\label{ctasinAniqui}
\end{equation}
when $t\rightarrow \infty$. Comparison with Eq.~(\ref{ctasinCoagu})
shows that at long times the concentration for the annihilation reaction is
exactly half that of the coagulation reaction.  This is in agreement with
the general result obtained above comparing the rates of change of the
concentrations.
The short-time behaviors for the two initial conditions are different, as
already evident in Fig.~\ref{fig:ctaniqui}.  For the random initial
condition an expansion of Eq.~(\ref{ctRandAniqui}) leads to
\begin{equation}
\frac{c(t)}{\lambda}=1-\frac{2\lambda
\sqrt{2 K_\alpha}}{\Gamma(1+\frac{\alpha}{2})} t^{\alpha/2}+\ldots
\label{ct1Rand}
\end{equation}
[compare with Eq.~(\ref{ct0Rand})].  For the periodic initial condition 
expansion of the denominator of Eq.~(\ref{cuAniquiPeriod}) for large $u$
and comparison with Eq.~(\ref{cu2}) leads to 
\begin{equation}
\label{ct1Perio}
\frac{c(t)}{\lambda}=1-
\frac{2}{\sqrt{\pi (2-\alpha)}}
\left(  \frac{\alpha}{2}\right)^{1/(\alpha-2)}
 z ^{1/(\alpha-2)}
\exp\left[-\frac{2-\alpha}{2} \left(  \frac{\alpha}{2}\right)^{\alpha/(2-\alpha)} z^{2/(2-\alpha)}
\right] +\ldots
 \end{equation}
where again  $z=[\lambda (2K_\alpha  t^{\alpha})^{1/2} ]^{-1}$ [compare
with Eq.~(\ref{ct0Perio})].

\section{Reactions in a lattice}
\label{lattice}
It is sometimes convenient not to carry out the continuous limit in
the formulation of the reaction kinetics for coagulation or
annihilation reactions.  Not only may the actual physical system be
discrete, but simulations usually involve discrete lattices, and finite
reactant size effects (i.e., small distance scale effects) are more
appropriately dealt with through discrete
formulations~\cite{ManAvr,KrebsEtAl,HenHin,AbaFriNic}.  Indeed, a discrete
viewpoint was the starting point of the continuum equations considered in
the previous section.  Furthermore, the discrete formulation allows 
consideration of different reaction rules such as
$A\emptyset A\rightarrow AAA$~\cite{HenHin}.

The hierarchy of differential-difference equations on a lattice of $L$
sites is
\begin{eqnarray}
I_0(t)&=& 1
\nonumber \\ \nonumber\\
\frac{dI_n}{dt}&=&a~_{0}D_{t}^{1-\alpha }\left[I_{n+1}(t)-2I_n(t)
+I_{n-1}(t)\right] \qquad 1\le n \le L-1
\label{IecuHierar}
\end{eqnarray}
for $n=1,2,\ldots$, where $a\equiv 2K_\alpha/(\Delta x)^2$,
$I_n(t)\equiv E(x_n=n\Delta x,t)$ for the coagulation reaction, and 
$I_n(t)\equiv r(x_n=n\Delta x,t)$ for the annihilation reaction.
The other boundary condition, at $n=L$, depends on which reaction is under
consideration and will be stated below.
The discrete form~\cite{AvrBurDoeJSP}
\begin{equation}
c(t)=\frac{1-I_1(t)}{\Delta x}
\label{ctdisc}
\end{equation}
was already introduced in Eq.~(\ref{c}).
From here on, we take $\Delta x=1$. 

\subsection{Coagulation in a segment}
The  problem  of the coagulation dynamics in a discrete segment
has been solved for the normal diffusive
case~\cite{ManAvr,KrebsEtAl,HenHin,AbaFriNic}.
Here we present the solution for the coagulation process of
subdiffusive particles on a chain of $L$ sites with periodic boundary
conditions, which is described by the hierarchy (\ref{IecuHierar}) together
with the additional boundary condition
\begin{equation}
I_L(t)= 0
\label{EecuHierarSegment}
\end{equation}
We solve this set of equations by means of the ansatz
\begin{equation}
I_n(t) = \sum_\omega b_n(\omega) E_\alpha(-a \omega t^\alpha)
\label{ansatz}
\end{equation}
which is closely related to the method of separation of variables
for the subdiffusive fractional differential
equation~\cite{MetKlaPhysRep,MetBarKlaPRL}. Again, $E_\alpha$ is the
Mittag-Leffler function of parameter $\alpha$.
Taking into account that 
\begin{equation}
\frac{d}{dt} E_\alpha(-\beta t^\alpha)
=-\beta~_{0}D_{t}^{1-\alpha }E_\alpha(-\beta t^\alpha),
\label{dEt}
 \end{equation}
the resulting eigenvalue problem can be solved in the same way as for
normal diffusion~\cite{HenHin,KrebsEtAl}. 
Notice that $E_\alpha(x)=e^{x}$ for $\alpha=1$ so that the
ansatz (\ref{ansatz}) takes the well-known form
for  normal diffusive particles (for which $\alpha=1$).
The  full solution is
\begin{equation}
I_n(t) = \left(1-\frac{n}{L}\right) +
\sum_{m=0}^{L-1} A_m \sin\left(m\frac{\pi}{L} (n-L) \right) E_\alpha(-a \omega_m t)
\label{solInt}
 \end{equation}
with
 \begin{equation}
\omega_m=2[1-\cos(m\pi/L)].
\label{omegam}
 \end{equation}
The coefficients $A_m$ are determined from the initial conditions.

For an initially random distribution of particles of concentration
$\lambda$ we have $I_n(0)=(1-\lambda)^n$ for $n=0,1,\cdots,L-1$, so that
\begin{equation}
A_m(t) = -\frac{2}{m\pi} \frac{(1-\lambda)^L m^2 \pi^2+(-1)^m L^2 \ln^2(1-\lambda)}
{ m^2 \pi^2+L^2 \ln^2(1-\lambda)}
\label{AmRandCoale}
\end{equation}
For an initially periodic distribution of particles separated by 
$1/\lambda$ lattice sites we have $I_n(0)=1-\lambda n$ for
$\lambda n\leq 1$ and $I_n(0)=0$ for $\lambda n\geq 1$. In this case
\begin{equation}
A_m(t) = (-1)^{m+1}  \frac{2 \lambda L \sin \left(m \pi/\lambda L \right)}
{ m^2 \pi^2}.
\label{AmPerioCoale}
\end{equation}

Figure~\ref{fig:ctcoagu} shows the  decay of the concentration $c(t)$ 
given by Eq.~(\ref{ctdisc}) with (\ref{solInt}) when the initial
distribution of subdiffusive particles on the lattice is random and when
it is periodic. At longer times both solutions approach those of the continuum
equations and would do so more clearly if we were to exhibit the solutions
to our difference equations for
larger lattices.  The marked deviation of the lattice solution from
the continuous asymptotic $t^{-\alpha/2}$ decay occurs when the
concentration on the lattice approaches its asymptotic 
(minimum) value of one particle on the entire lattice,
$c(t\rightarrow \infty)=1/L$.  This time can be estimated using the
continuum solution to occur at time $t_\times$ such that $c(t_\times)=1/L$.
In the figure this separation of solutions clearly does occur
when $c/\lambda = 1/\lambda L=0.1$, i.e., when $\log_{10}(c/\lambda)=-1.0$.

While the long-time behavior of the discrete and continuous solutions
coincide (independently of initial condition) until the reactant
concentration is
extremely low, the short-time behavior of the two solutions is expected to
differ, especially  when the initial distribution of particles is random.  
The difference has been established for particles undergoing normal
diffusion~\cite{AbaFriNic}.  To determine the short-time behavior on the
lattice we assume that the lattice is initially full, so that
$I_n(0)=0$ for $n\ge 1$. (The reasoning follows along the same lines for
an arbitrary concentration of particles, but
the equations are more cumbersome.)
Taking the time Laplace transform of Eq.~(\ref{IecuHierar}) gives 
\begin{equation}
\widetilde I_{n+1}(u)-(2+\frac{u^\alpha}{a})
\widetilde I_n(u)+\widetilde I_{n-1}(u)=0,
\qquad n \ge 1.
\end{equation}
This set of difference equations must be solved with the boundary
conditions $\widetilde I_0(u)=1/u$ and $\widetilde I_L(u)=0$.  Since we are
interested in the short-time (large-$u$) behavior, we can set
$L\rightarrow \infty$ without appreciably affecting the outcome.  With
these boundary conditions we essentially follow the procedure of Ref.
\cite{AbaFriNic} to find
\begin{equation}
\widetilde I_{n}(u)= \frac{1}{2^n u}
\left[ 2+ \frac{u^\alpha}{a} -
                  \left(\frac{u^{2 \alpha}}{a^2} +4 \frac{u^\alpha}{a}
 \right)^{1/2}
 \right]^n.
 \end{equation}
In particular, $\widetilde I_1(u)=a/ u^{1+\alpha}+\ldots $
for $u\rightarrow \infty$,
so that with the help of appropriate Tauberian theorems we find that
$I_1(t)=at^\alpha/\Gamma(1+\alpha)+\ldots $ for $t\rightarrow 0$.
Using Eq.\ (\ref{ctdisc}), we calculate the short-time behavior of the
reactant concentration
\begin{equation}
\frac{c(t)}{\lambda} =
1-\frac{2  K_\alpha \lambda^2}{\Gamma(1+\alpha)} t^\alpha +\ldots.
\end{equation}
Note that with an initially full lattice there is no distinction between an
initially random and initially periodic distribution of particles.
For an arbitrary random initial concentration one still finds that
$c(t)/\lambda-1\sim t^{\alpha}$. Comparison with Eq.~(\ref{ct0Rand}) shows
that at short times, as in the case of normal diffusion~\cite{AbaFriNic}, the 
concentration on a lattice decays more slowly than in the continuum.  This
is not easily visible on the scale of Fig.~\ref{fig:ctcoagu}.

\subsection{Annihilation in a segment}
Now consider the annihilation reaction $A+A\rightarrow 0$ and suppose that
initially there are $N$ particles on a segment of length $L$.  The
hierarchy (\ref{IecuHierar}) is now augmented with the second boundary
condition
\begin{equation}
 \label{recuHierarSegment}
   I_L(t)=\begin{cases}
1, & N=\text{even}\\[9pt]
0, & N=\text{odd}
\end{cases}
\end{equation}
where we note that now $I_n(t)\equiv r(x_n=n\Delta x,t)$. 
This condition is of course a result of the fact that the reaction does not
change the parity of the number of surviving particles. We proceed as in the
coagulation problem and write the solution as
\begin{equation}
I_n(t) = I^*_n +
\sum_{m=0}^{L-1} A_m \sin\left(m\frac{\pi}{L} (n-L) \right) E_\alpha(-a \omega_m t)
\label{solIntAniqui}
 \end{equation}
where 
\begin{equation}  
I^*_n=I_n(\infty)=\begin{cases}
1, & N=\text{even},\\[9pt]
1-\frac{n}{L}, & N=\text{odd},
\end{cases}
\label{I*def}
 \end{equation} 
$\omega_m$ is given in Eq.\ (\ref{omegam}), and the $A_m$ are coefficients
determined from the initial conditions.

For a random initial condition of particle concentration $\lambda$ we have
$I_n(0)=\frac{1}{2}+\frac{1}{2}(1-2\lambda)^n$ for $n=0,1,\cdots,L-1$,
so that
 \begin{equation}
A_m(t) = \frac{1}{m\pi} 
\frac{\left[(-1)^N-(1-2\lambda)^L\right] m^2 \pi^2+
				\left[1-(-1)^{m+N}\right] L^2 \ln^2(1-2\lambda)}
{ m^2 \pi^2+L^2 \ln^2(1-2\lambda)}.
\label{AmRandAniqui}
 \end{equation}
For a periodic initial distribution with particle separation $1/\lambda$ we
have
\begin{equation}
I_n(0)=
\begin{cases}
2j+1-\lambda n & 2j/\lambda\leq n\leq (2j+1)/\lambda \\[9pt] 
-2j-1+\lambda n & (2j+1)/\lambda \leq n\leq (2j+2)/\lambda
\end{cases}
\end{equation}
which leads to 
 \begin{equation}
A_m(t) = (-1)^N \left[1-(-1)^N(-1)^m\right]^2  \frac{\lambda L}{m^2 \pi^2}\tan
\left(\frac{m \pi}{2\lambda L}\right).
\label{AmPerioAniqui}
 \end{equation}

Figure~\ref{fig:ctaniqui} shows the evolution of the concentration
as given by Eq.~(\ref{ctdisc}) with (\ref{solIntAniqui}) 
for a random initial distribution (circles) and a periodic initial
distribution (squares).  A number of points are noteworthy.  First, the
solution of course decays to a finite constant (one particle remaining)
when there is initially an odd number of particles in the system, and to  
zero when the initial number of particles is even.  The deviation of the
solutions on a finite segment from those in an infinite continuum again set
in around the time when $c=1/L$ (strictly speaking around $2/L$ for the
even case).  Note that in the continuum solution one does not distinguish
between the even and odd particle number cases: the boundary condition 
(\ref{rccxinfini}) is an average of the two, and so the continuum curves
should be compared to the average of the discrete ones for the two boundary
conditions.

\section{Conclusions}
\label{conclusions}

We have considered the coagulation dynamics $A+A\rightarrow A$
and the annihilation dynamics $A+A \rightarrow 0$
for particles moving subdiffusively in one dimension.  This scenario
combines the ``anomalous kinetics" and ``anomalous diffusion" problems,
each of which leads to interesting dynamics separately and to even more
interesting dynamics in combination.  The fractional diffusion equation
plays a central role in our analysis and allows the exact calculation of
the density $c(t)$ within this formulation.
We have calculated $c(t)$ explicitly for all times for the coagulation
reaction in a one-dimensional continuum for two initial distributions,
a random (Poisson) distribution and a periodic distribution. 
This calculation is based on the empty interval method.
Using the odd/even interval formulation, we have also obtained
an explicit solution for the annihilation reaction with a random initial
distribution.  For a periodic initial distribution we are only able to
calculate the Laplace transform of $c(t)$ analytically and must perform the
inversion numerically.

For the coagulation reaction the empty interval method also leads to an
explicit solution for the interparticle distribution function
$p(x,t)$.  At long times we find a universal expression for this
distribution (i.e., one independent of initial distribution), in terms of
a single scaled variable.  Anomalous diffusion is characterized by
the exponent $\alpha$ 
introduced in Eq.~(\ref{meansquaredispl}), ordinary diffusion
corresponding to $\alpha=1$. 
Deviations from ordinary diffusion lead to
a curious interplay.  On the
one hand, with decreasing
$\alpha$ (and hence increasingly subdiffusive
motion) the decay of the particle density towards extinction becomes
increasingly slower and in this sense increasingly
different from law of mass action behavior.  On the other hand, 
for the coagulation reaction the
spatial distribution of initially randomly distributed reactants
remains more Poissonian for all time as $\alpha$ decreases; indeed,
as $\alpha$ deviates from unity the relatively empty regions
around each particle
that are tantamount to (and indeed explain) anomalous kinetics in the
usual diffusion-limited case become more populated.

We have also considered the problem of subdiffusion-limited reaction on
discrete lattices.  We have presented a hierarchy of
differential-difference equations for
the empty interval probability $I_n(t)\equiv E(x_n,t)$ and the even/odd
interval probability $I_n(t)\equiv r(x_n,t)$, and have expressed the
solutions
$I_n(t)$ as a superposition of subdiffusive modes whose decay is governed by
the Mittag-Leffler function in the same way that the exponential function
governs the decay of ordinary diffusive modes.  As in ordinary
diffusion~\cite{AbaFriNic}, the associated
concentrations $c(t)$ differ from the continuum solutions at early times.
At long times the lattice solutions deviate from those of the continuum
systems when there is only a small number (one or two) of particles
remaining in the system.

\section{Acknowledgments}

This work has been supported in part by the Ministerio de Ciencia
y Tecnolog\'{\i}a  (Spain)
through Grant No. BFM2001-0718 and by the Engineering Research Program of
the Office of Basic Energy Sciences at the U. S. Department of Energy
under Grant No. DE-FG03-86ER13606.  SBY is also grateful to the DGES
(Spain) for a sabbatical grant (No.\ PR2000-0116) and to the
Department of Chemistry and Biochemistry of the University of
California, San Diego for its hospitality.

\end{document}